\documentclass[12pt]{iopart}
\usepackage{graphicx}

\begin{document}
\title{Opinion dynamics on a group structured adaptive network}
\author{Floriana Gargiulo}
\address{LISC, Cemagref, 24 Avenue de Landais, Clermont Ferrand, France}
\ead{floriana.gargiulo@cemagref.fr}
\author{Sylvie Huet}
\address{LISC, Cemagref, 24 Avenue de Landais, Clermont Ferrand, France}
\ead{sylvie.huet@cemagref.fr}
\begin{abstract}
Many models have been proposed to analyze the evolution of opinion structure due to the interaction of individuals in their social environment. Such models analyze the spreading of ideas both in completely interacting backgrounds and on social networks, where each person has a finite set of interlocutors. Moreover, the investigation on the topological structure of social networks has been the object of several analysis, both from the theoretical and the empirical point of view. In this framework a particularly important area of study involves the community structure inside social networks.
In this paper we analyze the reciprocal feedback between the opinions of the individuals and the structure of the interpersonal relationships at the group level. For this purpose we define a group based random network and we study how this structure co-evolves with opinion dynamics processes.
We observe that the adaptive network structure affects the opinion dynamics process helping the consensus formation. The results also show interesting behaviors also in regards to the size distribution of the groups and their correlation with opinion structure.
\end{abstract}
\section{Introduction}
In the last decades many different reaction processes on complex networks have been intensively studied \cite{pastor2004evolution}: from epidemics \cite{pastor2001epidemic}, \cite{pastor2002immunization}, to malware diffusion in electronic technology \cite{hu2009wifi}, collective behaviors \cite{helbing2000simulating}, innovation diffusion and opinion dynamics \cite{castellano2009statistical}, \cite{galam2008sociophysics}. 

Opinion dynamics, in particular, is a symmetric contact process, where people can influence each other and induce other people to change mind on some particular topic. Many models have been proposed to study the spreading of opinion: some of these models describe opinion as a discrete Boolean choice, like the Voter model \cite{clifford1973model}, \cite{holley1978survival} or the Sznajd model \cite{stauffer2002sociophysics}. These formalizations can describe, for example, the positions on elections in majoritarian systems (where only two parties are present). Other models take into account the fact that, for some kinds of situation, people can have a certain continuous level of agreement on a topic, like for example regarding the involvement of a country in a war,  the production of nuclear energy, the choice of organic food. 

The first model describing continuous opinion interaction, also introducing the concept of bounded confidence (people with  too different opinions among themselves cannot influence each other) is known as Deffuant (or \emph{bounded confidence}) model \cite{deffuant2000mixing}. Some different implementations of this model taking into account a rejection process \cite{huet2008rejection} or a different type of tolerance threshold connected to the opinion \cite{gargiulo2008can}  have been proposed in the last years.

The interest toward opinion dynamics increases coupling these phenomena with the investigation on the topological structure of social networks. Recently this topic has been the object of many analysis, both from the theoretical and the empirical point of view (using for example the WEB 2.0 technologies).

Different kinds of network topologies have been tested both to prove the robustness of the opinion dynamics models and for identifying preferential channels of opinion spreading \cite{ard2004role}. At the beginning all the considered network topologies were static: namely, the connections among the persons did not vary in time. This approximation is reasonable if we consider that the processes happening on network (like opinion spreading in this case) have a different, much shorter, time scale than the process that changes the structure of network (rewiring mechanism, cutting of links).  Incidentally many works have recently been done regarding evolving network topology and their adaptation to the social background \cite{gross2008adaptive}: as people can influence each other to induce a change of mind, the difference of opinion on some very important topics can also lead to a group change and the breaking of some social contacts. In other terms, since people prefer to be surrounded by persons sharing similar opinion (homophily), it is quite likely that the change of opinions due to the opinion dynamics processes can lead to the change of the network structure. An interesting analysis of the co-evolution of opinions and networks is presented in \cite{kozma2008consensus}. 

Regarding social network analysis, sociologist and network scientists agree on the fact that social networks present sometimes community based structures: analyzing networks at different scales, it is possible to identify groups of persons much more interconnected among them than with the rest of society \cite{girvan2002community}. Many different algorithms have been created to identify communities on large networks \cite{2009community} and many models have been proposed to explain the mechanism leading to the formation of such underlying structures \cite{palla2007quantifying}.

From another point of view, such an approach is widely justified by the classical approaches in social-psychology which consider that an individual can change groups in case of disagreement with his own group. This aspect is developed later, at the same time the model is presented.

In this paper we argue that the community structures on network - or group - have a strong connection to the opinion of the agents and we propose a model of co-evolution of the opinions and of the group structures present in the social network.

In section \ref{MD}, the details of the model are argued and explained; in section \ref{SP}, the details of the simulation and the parameter choices are explained and finally, in section \ref{res}, the results are presented. The result section is divided into two parts: the first one analyzes the global effect of the adaptive network structure in an opinion dynamics process. It is observed that the mechanism introduced favors the creation of consensus, lowering the typical Deffuant model transition threshold to consensus. In the second part the structure of the communities, at the end of the simulation is analyzed. We observe that in some cases, in the pluralistic phase of the system, the community sizes present a very heterogeneous distribution and we argue that the size of the groups has a strong correlation with  the opinions of the agents.

\section{Model description}\label{MD}
Others opinion is a source of cognitive inconsistency! That is what Festinger \cite{festinger1957Dissonance} argued adding that it is experienced as dissonance. According to him, the dissonance is a psychological discomfort or an aversive drive state that people are motivated to reduce, just as they are motivated to reduce hunger. In his balance theory, \cite{heider1946} used a similar concept and called it imbalance. More recently, \cite{MatzWood2005} showed that, as the dissonance and balance theories suggest, the disagreement from others in a group produces cognitive inconsistency and the negative states of dissonance or imbalance.

The groups are a privileged place of interaction between people and the exchange with others can lead to dissonance. They are thus at the same time the entity creating dissonance and the one reducing it. Indeed, three strategies can be chosen to reduce its dissonance created by the heterogeneity of the opinion inside its group: changing its own opinion to agree with others in the group, influencing others to change their opinions, or joining a different, attitudinally more congenial group. The three ones reduce dissonance \cite{MatzWood2005}. 

The two first relates to the individual interactions which are often based on similarity and have been extensively studied in the attraction paradigm \cite{Byrne1971} and other theories on interpersonal interactions as the social judgment theory \cite{SherifHovland1961}. The third can be linked to the membership of individuals into a group and, therefore, to their position of the individual  in the social network. 

The model presented in this paper aims at describing simultaneously these two levels (the individuals and the network)  and therefore can be divided into two sub-modules: the first one deals with opinion dynamic between two individuals (ODM), and the second one deals with  the choice of a single individual of being a member of a determined group (MDM). The last module, in particular, describes how a person shapes its relationship scenario according to its opinion.

The effects of combining these two kinds of dynamics is equivalent to study an opinion dynamics model on an evolving social network that step by step adapts itself to the opinion structure of the society. We start therefore to describe the initialization of the network and how the group structures are defined. After the initialization the two dynamical modules are described.

\subsection{A group-based network structure}

A very common problem in network science is finding community structure in large networks. In this paper we start from a different  point of view: we generate a network with an already well defined community structure and we study the evolution of such communities in time. For the moment we will not consider the possibility of overlapping communities.
The idea is to initialize the model with a random structure, but with the possibility of clearly identifying group structures.
We start fixing the initial number of communities present in the model ($G_0$).
At the moment of the initialization each agent is selected and is randomly attributed one of the $G_0$ groups. After all  agents have been located into the group structures, the links are created:

\begin {itemize}
\item All the agents in the same group are linked together
\item All the pairs of agents that are not members of the same group are connected with a probability $p_{ext}$.
\end {itemize}

Using such a procedure, each group, at the beginning contains a number of members $n_0$ such that: $\langle n_0\rangle=N/G_0$, where $N$ is the total number of agents. The variance is very small so that we can assume $\langle n_0^2\rangle\sim \langle n_0\rangle^2$.
Also the degree distribution is Poissonian and the average degree is given by: $<k_0>=(N/G_0-1)+p_{ext}(N-N/G_0)$.
The average number of in-group connection in a community is given by $\langle L_{in}\rangle=\langle n_0(n_0-1)/2\rangle$.
The average number of out-group connections for each community is given by: $\langle L_{out}\rangle=\langle n_0p_{ext}(N-n_0)\rangle$.

If we want to initialize the network in order to have a well defined topological community, the number of internal connections, $L_{in}$, should be bigger than the number of out-group connections, $L_{out}$, (according to the basic community description presented for example in \cite{radicchi2004defining}). This condition reflects on a constraint connecting $G_0$ and the out-group connectivity $p_{ext}$:
\begin{equation}
\frac{1}{2}\left(\frac{N}{G_0}-1\right)> p_{ext}N\left(1-\frac{1}{G_0}\right)
\end{equation}

The initial network structure is opinion independent: after the topology initialization, an opinion $\vartheta_i(t=0)$, continuously ranging in the interval [0,1], is randomly assigned to each agent. Therefore at the beginning of the simulation all the groups will have an average opinion $O_I=\langle\vartheta\rangle_I\sim 0.5$.

After the initialization, at each time step, all the agents are updated in a random order: with a probability $p_{change}$ an agent changes group (applying MDM), while with the probability $1-p_{change}$ it performs opinion dynamics (applying ODM).

From the point of view of the groups, therefore, the only possible actions are "gaining  a member" and "losing a member". The mechanisms of group merging and splitting are not considered in this model and therefore the number of groups remains constant. It is also important to observe  that, with the evolution of the system, it can happen that some groups loose the typical characterization of the community (a higher number of internal than of external links). Therefore, in the following, we will always adopt the term \emph{group} for indicating these macro--structures independently from the fact that they are or not communities, from the topological point of view.

\subsection{Opinion dynamics module (ODM)}

A selected agent can change its opinion with the interaction with an other agent in its neighborhood, interacting both with someone in its own group and with someone connected with it through external links.
The agent, therefore, randomly selects the second agent of the opinion dynamics process between the whole set of its connections.
The two agents interact according to the Deffuant model \cite{deffuant2000mixing}:
\begin{equation}\label{deffuant}
\mbox{if }|\vartheta_i(t)-\vartheta_j(t)|<\varepsilon \qquad 
 \vartheta_i(t+1)= \vartheta_i(t)+\mu(\vartheta_j(t)-\vartheta_i(t))
\end{equation}
the same for $j$. In the equation, $\vartheta_i$ is the opinion of agent $i$.

Two parameters describe completely this model:

\begin{itemize}
\item the bound confidence, $\varepsilon$, describes the tolerance of the agents, namely the maximum distance between opinions allowing them to influence each other.
\item the reciprocal shift, $\mu$, that describes how much the two agents approach after the interaction. In general we will fix $\mu=0.5$.
\end{itemize}

The Deffuant model is very robust on different types of static topologies: on complete graphs, like on random networks, lattices, and on scale free structures, it presents a phase transition for $\varepsilon=0.5$ \cite{fortunato2004universality}. Two different regimes can be identified: for  $\varepsilon<0.5$ at the end of the simulation many different opinion clusters remain; for $\varepsilon>0.5$ the consensus is reached (a single opinion cluster in the center of the opinion space).

\subsection{Membership dynamics module (MDM)}

People can change their opinion with interactions, but they can also decide to change their group membership if their opinion is too different from the opinions of the other agents in the group. With this action an agent preserves its opinion but it changes the topological structure of the social network and the average opinions of the groups.\\
Also for the process of changing group the same bound for tolerance introduced in opinion dynamics is used: an agent, $i$, member of the group $I$ changes group only if:

\begin{equation}\label{cg}
    |\vartheta_i(t)-O_I(t)|>\varepsilon
\end{equation}
where $O_I(t)=\langle\vartheta(t)\rangle_I$ is the average opinion of group $I$.

If this condition is realized the agent selects a new group, $J$, between the groups where it has an external link, with probability:
\begin{equation}\label{pselgroup}
 P_{i\rightarrow J}=\frac{1-|\vartheta_i(t)-O_J(t)|}{\sum_{J\supset j\in \mathcal{V}(i)}\left(1-|\vartheta_i(t)-O_J(t)|\right)} 
\end{equation}
where $\mathcal{V}(i)$  is the neighbourhood of node $i$.\\

After the individual has changed groups, all its links are rewired:
\begin{itemize}
  \item All the previous links are canceled
  \item The agent is connected with all the people of the new group
  \item The agent is connected to agents outside its group with probability $p_{ext}$
\end{itemize}

The choice of rewiring also the external links is mostly due to the fact that, if such procedure is not performed, at some point of the simulation, all the external links can become internal and a group can remain isolated. Moreover, since also the external links are subject to change in time, instead of introducing a second rewiring procedure relative to external links, we group together the two rewiring processes. 

\section{Simulation parameters}\label{SP}
The model is completely described by five independent parameters: the two parameters of the bounded confidence model ($\varepsilon, \mu$), the initial number of groups ($G_0$), the probability of having an out--group connection ($p_{ext}$) and the probability of choosing opinion dynamics or group dynamics ($p_{change}$).
A priori, the parameter appearing in eq. \ref{cg} could be considered different from the $\varepsilon$ of bounded confidence, but, since both the parameters are related to the \emph{tolerance} of an individual to others' opinions we decided to consider these two parameters equal. Therefore, a complete exploration of the system represents quite a hard target. 

Based on what we know from the Deffuant model behaviour, we  will focus these first analysis on some regimes where the dynamics should be more interesting. We will be mainly interesting in varying $\varepsilon$ that we know as the main determinant of the Deffuant model behaviour. For the rest, we will perform our analysis on a system composed by $N=5000$ agents. We will fix the probability for an external connection to $p_{ext}=0.001$. In the first part of the analysis, we will consider an initial number of groups $G_0=500$. It means that, on average, at the beginning of the simulation each agent will have $k_{in}\sim 9$ in-group links and $k_{out}=5$ out-group links. The attraction parameter for the Deffuant BC model will be fixed for all the simulation to $\mu=0.5$, assuming therefore that, after a successful interaction, two agents will share the same opinion. We will always consider that the ODM and the MDM procedures happen with the same probability, fixing $p_{change}=0.5$. Nevertheless, we will compare some result of the simulation with the extreme case where the membership dynamics module is not applied ($p_{change}=0$), restoring the usual BC model on a group based static network. In any case, the results are calculated as the average of 100 independent stochastic realizations.

The system is evolved until an equilibrium situation is reached, where neither the opinions nor the topology is going to change anymore. This stable point is reached when  two simultaneous conditions are reached: all the opinions of the connected agents are either equal or differ more than the threshold and the opinion distance of each agent from the average of its group is smaller than the opinion distance with any of its external contacts.

The result section is divided into two different sections: the first one describes the results at a global level and the general critical properties of this model with respect to the static version of Deffuant model. In the second part, the local properties of the system and the group structure are analyzed. For this first analysis, we will focus our study on the particular behaviour obtained for $\varepsilon=0.1$.

\section{Critical behavior}\label{res}

It is well known that the Deffuant model presents two possible equilibrium situations: a consensus scenario where all the opinions of the agents converge to a central cluster, or a pluralistic scenario where the equilibrium is given by different opinion clusters. The transition between the two situations can be represented as a phase transition with the critical parameter $\varepsilon$. The critical value where the transition happens, for the Deffuant model on different static topologies is $\varepsilon_c=0.5$: for $\varepsilon<\varepsilon_c$ the system stabilizes in a pluralistic configuration, while, on the opposite, for $\varepsilon\geq\varepsilon_c$, a consensus state is reached. \\For the following discussion it is important to stress out the difference between the concept of group, that is a topological attribute that we defined on the network, and the concept of opinion cluster, that has nothing to do with topology. According to usual definitions, two agents are considered members of the same cluster if their opinions differ less than a fixed threshold ($cl_{thres}=0.01$).

\begin{figure}[!h]
  \includegraphics[width=12cm,clip=true]{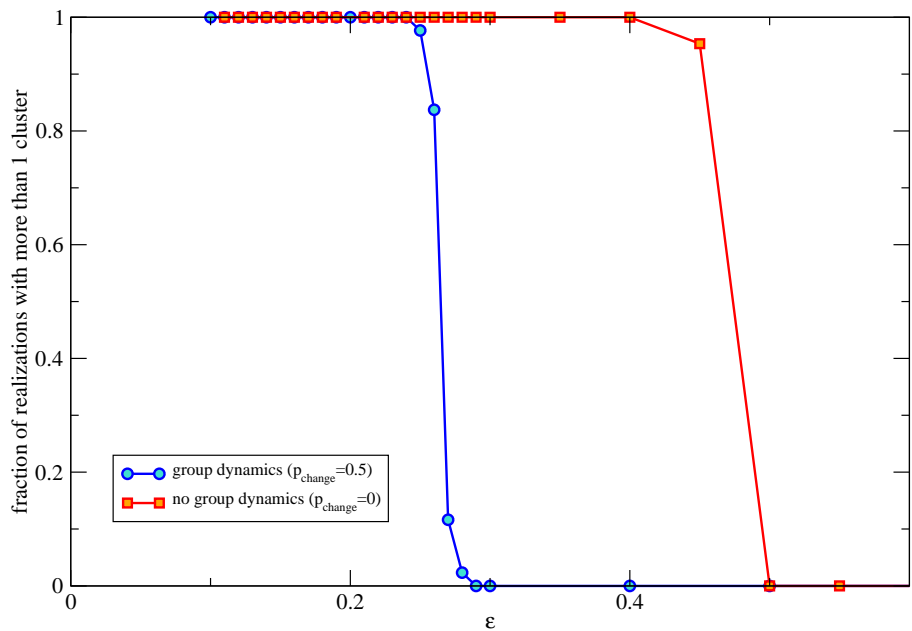}
  \caption{ Fraction of non-consensus realizations (more than one opinion cluster) as a function of $\varepsilon$ on 100 replicas. The red line refers to the case where no opinion dynamics is performed ($p_{change}=0$)}\label{fig_ncl}
\end{figure}

In Figure \ref{fig_ncl}, the fraction of realizations that end up with a single opinion cluster as a function of $\varepsilon$ is analyzed. 
As we can see, the dynamical adaptive network structure changes the critical properties, lowering the threshold value for $\varepsilon$; the line  describing groups dynamics, in fact, goes to zero much more rapidly than the other one showing  the role of the network co--evolution in favoring the consensus formation. The transition threshold, for adaptive networks, stabilizes around $\varepsilon_c\sim 0.3$.\\

To illustrate and better understand the observed phenomenon, we're presenting the trajectory of individual opinions and group opinions for one replicate for $\varepsilon=0.3$. We use two kinds of graphs where the abscissa corresponds to the group number while the ordinate corresponds to the opinion. The upper graphs shows the evolution over time of opinion of the individuals. Each dot is an individual position with on abscissa its group number, and, on ordinate, its opinion. The lower graphs shows the same evolution over time for the groups. Each circle represents a group with its number on the abscissa and its opinion (i.e. the average one of its member) on ordinate. The surface of the circle indicates the relative size of the group compared to the population size.

\begin{figure}[!h] 
  \includegraphics[width=12cm,clip=true]{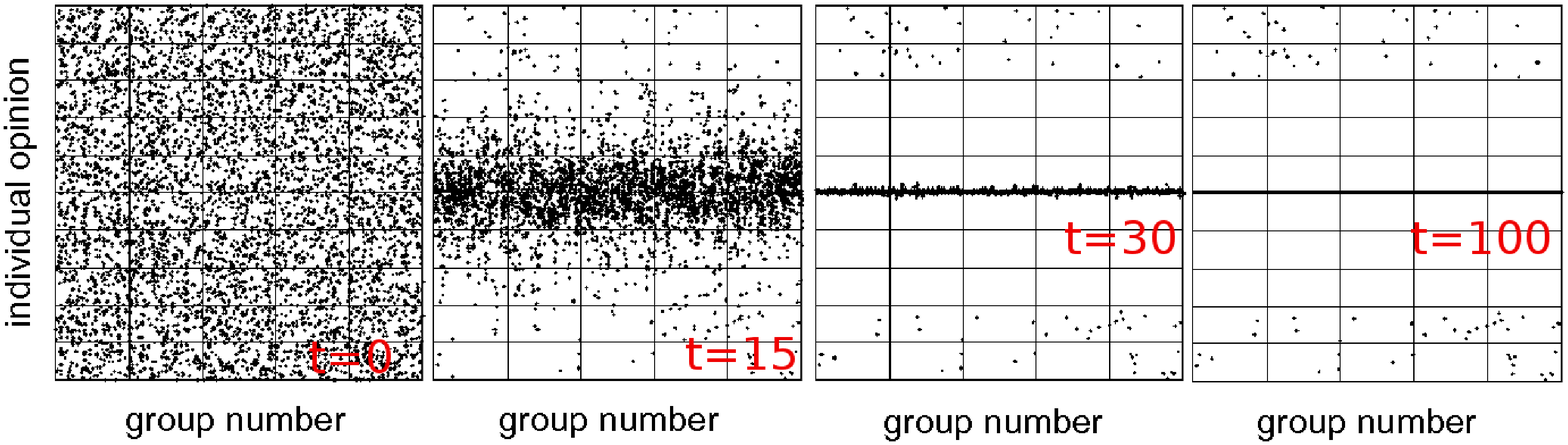}
  \includegraphics[width=12cm,clip=true]{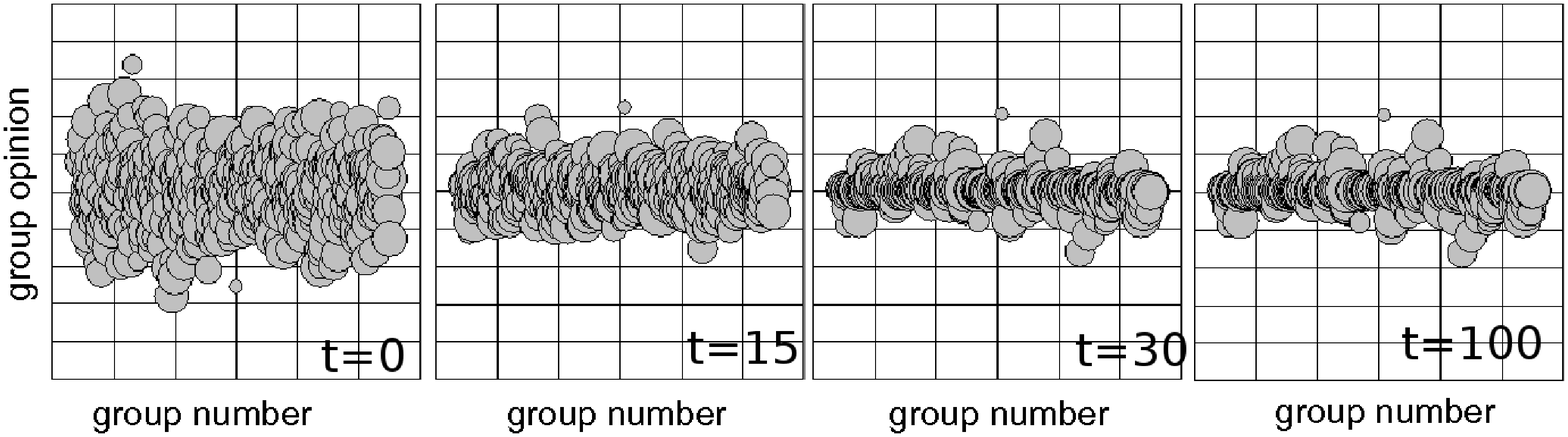}
  \caption{ Evolution of a population in one realization for $p_{change}=0$ and $\varepsilon=0.3$ (from the left to the right, after 0, 15, 30, 100 iterations). Upper plots: individual opinions. Lower plots: average group opinions and relative group size (radius of the ball) }\label{fig_1real_BCalone}
\end{figure}
\begin{figure}[!h]
  \includegraphics[width=12cm,clip=true]{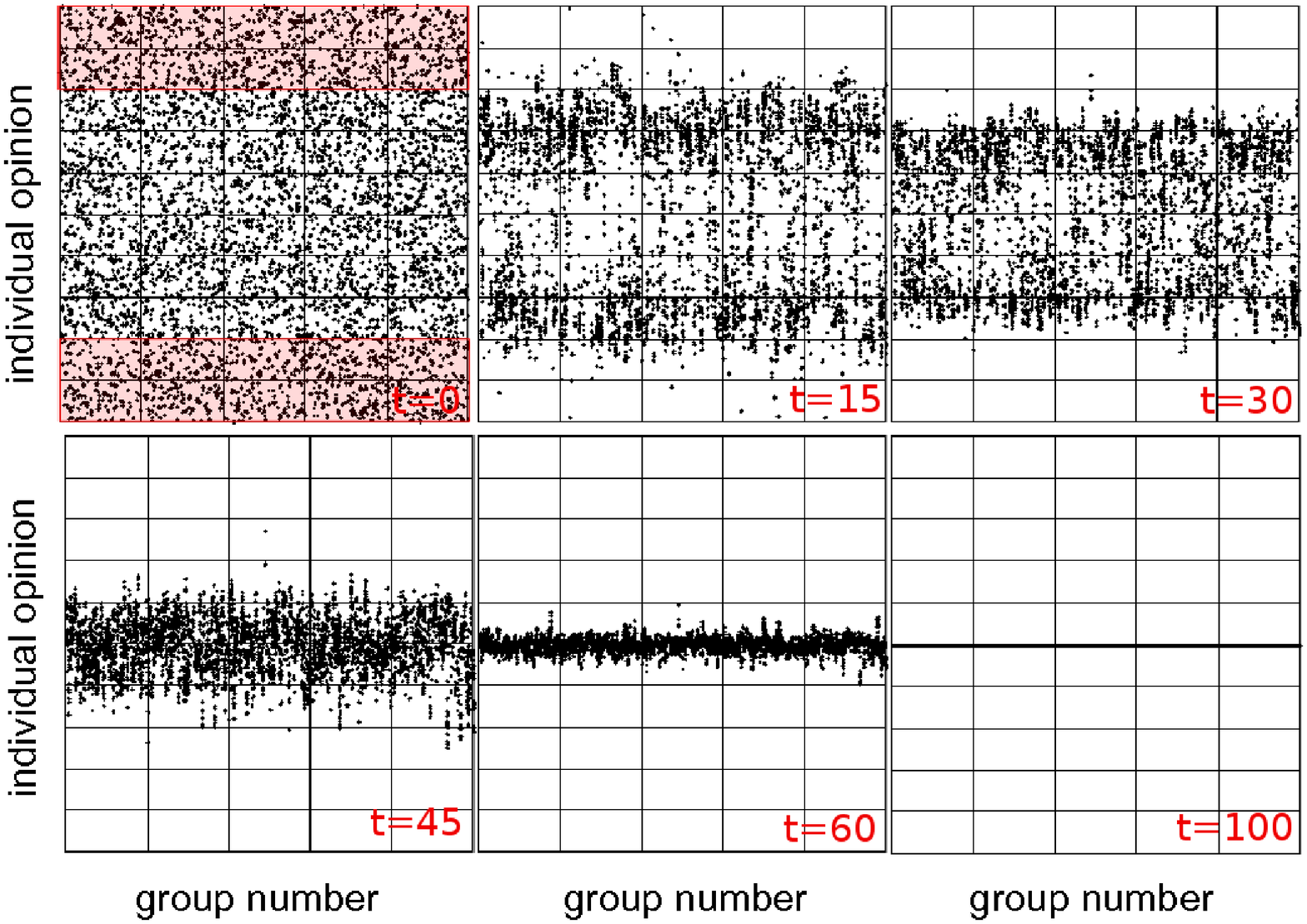}
  \includegraphics[width=12cm,clip=true]{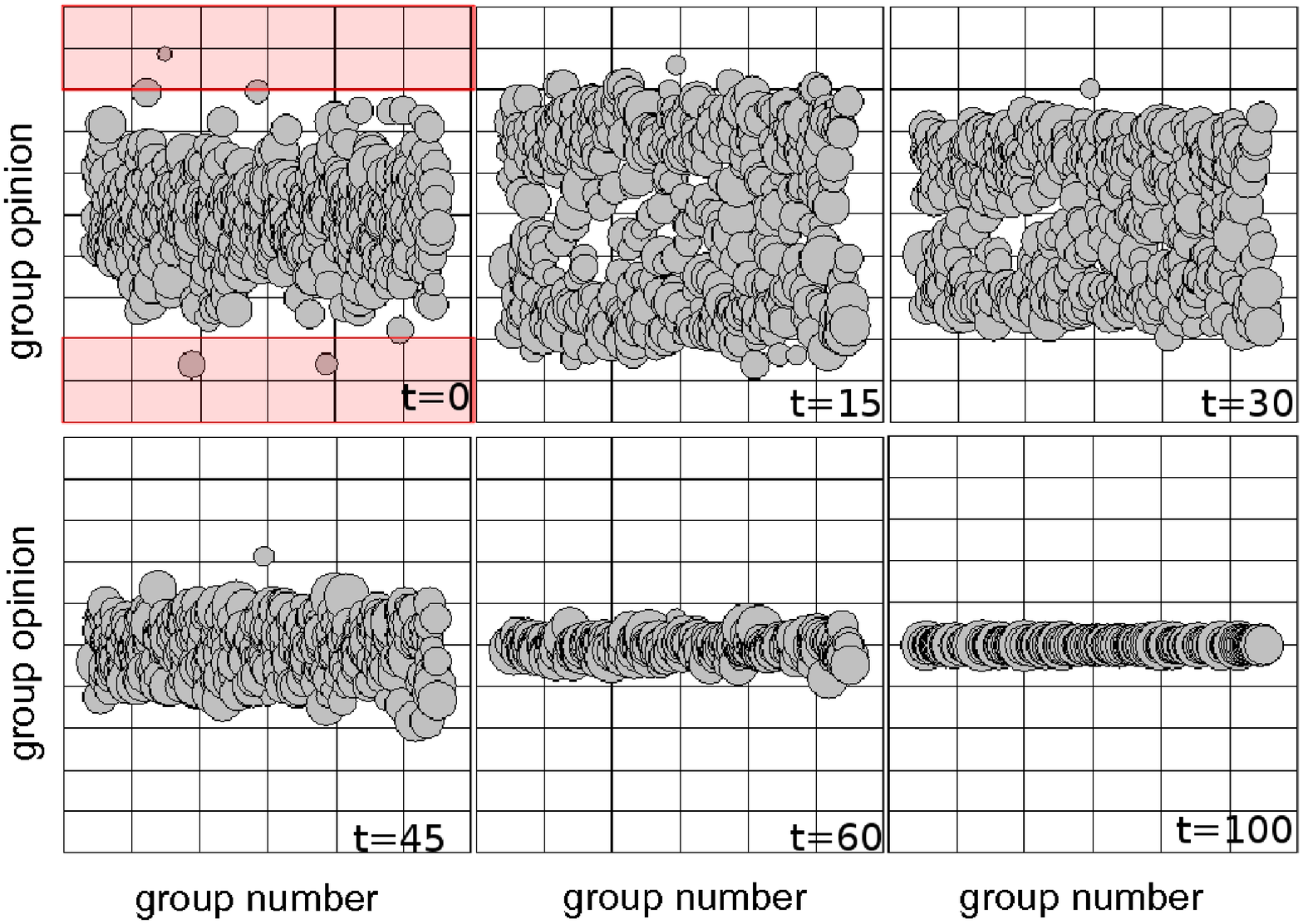}
  \caption{Evolution of a population in one realization for $p_{change}=0.5$ and $\varepsilon=0.3$ (from the left to the right, after 0, 15, 30, 45, 60, 100 iterations)
. Upper plots: individual opinions. Lower plots: average group opinions and relative group size (radius of the ball) }\label{fig_1real_BCgroups}
\end{figure}

Figure \ref{fig_1real_BCalone} shows the evolution of a population having only the Deffuant opinion dynamic module ($p_{change}=0$). You can see that, as we know from Ben Naim (\cite{ben2003bifurcations}), Lorenz (\cite{lorenz2007continuous}) and Laguna (\cite{laguna2004minorities}), the convergence happens very quickly and some individuals remain on the border of the attitude space. They have been "forgotten" by the others due to the speed of the dynamics which is $\mu = 0.5$. These "forgotten" extremists are usually called "minor clusters" in the literature.

Figure \ref{fig_1real_BCgroups} shows an evolution of a population submitted to the coupled opinion and group dynamics. You can observe that the convergence is slower than in the latter case and allows the population not to "forget" someone. Then the consensus after 100 steps is total and all individuals have a centred opinion.
\begin{figure}[!h]
  \includegraphics[width=12cm,clip=true]{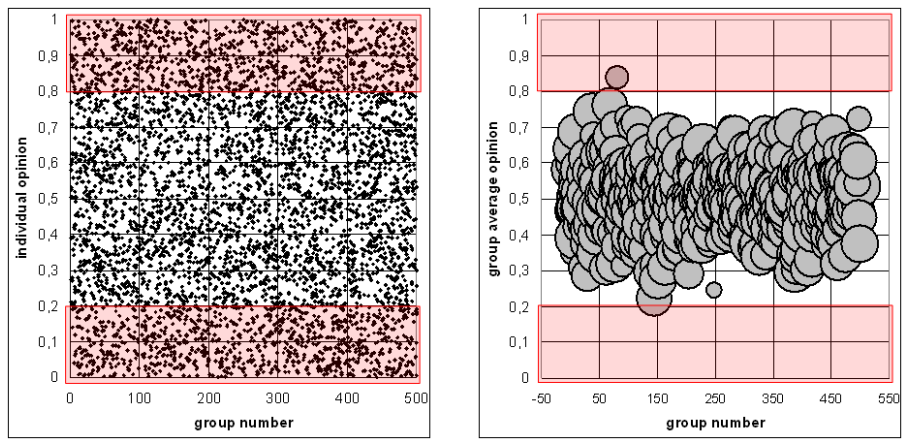}
  \caption{Areas where individuals are susceptible to change group at the beginning of simulation for $\varepsilon=0.3$.}\label{fig_areas03}
\end{figure}

Figure \ref{fig_areas03} better explains how nobody can be "forgotten" by the dynamics. Considering that the theoretical average opinion of a group is $O_I\approx0.5$, we can deduce that, theoretically, all people situated in the shadow rectangles are susceptible to change groups because they are far from the average opinion of their group. Practically, you can see on the right that, from the beginning, the stochasticity of the model implies the average is not totally 0.5. Then, there is always a group whose the average opinion is close enough to allow to extremists to join the other people in their convergence to the centre. That is the reason why minor clusters do not exist in this "group" version of the Deffuant model. 

The fact that  minor clusters do not  appear has a strong impact on the critical behaviour of the model: as Ben Naim pointed out in \cite{ben2003bifurcations}, Deffuant model for opinion dynamics presents 4 types of bifurcations at different values of $\varepsilon$:  separation of two minor clusters symmetrically from the central one at $\varepsilon\sim0.5$ (transition from consensus to pluralism); creation of two major clusters from the central one at $\varepsilon\sim 0.266$; separation of a minor central cluster at $\varepsilon\sim0.222$; growth of the central cluster and shift to extremist positions of the two side clusters $\varepsilon\sim0.182$. 

The model we present here is the usual Deffuant model, but applied on a group--based, dynamic network. Therefore we shod expect to observe a critical behaviour similar to the Deffuant model on static topologies. As we observed before the effect of group dynamics is the suppression of minor clusters. Therefore we can argue that the critical transition observed for static topologies at $\varepsilon\sim0.266$, between 2 non central clusters and 1 central clusters plus minor structure, directly corresponds, in our case, to the transition between 2 non central clusters and consensus.
A more detailed analysis of such behaviour is given in \cite{Gargiulo:WEIN}.

It can be interesting to approach this matter from the behaviour of the Hegselman and Krause model \cite{hegselmann2002opinion} which is also a bounded confidence model with a pseudo-group approach. Indeed, the group of an individual is dynamic in this instance and corresponds to all the individuals situated at an opinion distance around the individual of almost $\epsilon$. The individual interacting with its group adopts its average opinion. This Hegselman and Krause version of the bounded confidence model also does not exhibit some minor clusters.

\section{Detailed results in a pluralistic scenario}

In this section we are going to analyze the network and group structure together with their relation with the opinion distribution, for fixed values of $\varepsilon$ in the pluralistic region.
The first quantity we are going to analyze is the group size distribution.
\begin{center}
\begin{figure}[!h]
  \includegraphics[width=10cm,clip=true]{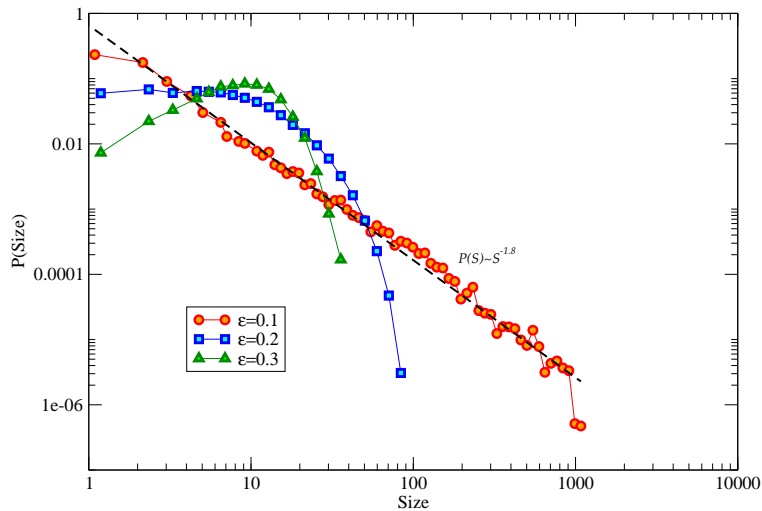}
  \caption{Size distribution for groups for $\varepsilon=0.1,0.2$ and $0.3$.The result is averaged on 100 replicas of the simulation.}\label{fig_psize}
\end{figure}
\end{center}

The figure \ref{fig_psize} presents the size distribution of groups for different values of $\varepsilon$ (0.1, 0.2 and 0.3). For the largest values of $\varepsilon$ (0.2 and 0.3), the distribution of size remains close to the initial one.

However, we can see on the same figure \ref{fig_psize} than the size distribution of the groups, for $\varepsilon=0.1$  follows a  clear power-law distribution with exponent $\eta=-1.8$. In this case, therefore, the group size shows a strong heterogeneity. In particular, we can observe than almost the half of the initial groups tend to disappear (to contain a single element at the end of simulation), while a small number of extremely populated groups appear. 

The birth of heterogeneity is due to a preferential attachment mechanism, hidden in the group changing process. In fact, an agent, deciding change groups, looks around to the groups it can retrieve direct information on, namely, the groups to which it is externally connected. If a group is bigger, the probability that some external connections end-up there is higher and therefore it is higher the probability to be selected as a possible destination. If we allow the agents to choose the group with a rational choice, namely, selecting between all the possible groups, the heterogeneity is not observed.

This strong heterogeneity for the group size tends to disappear for higher values of $\epsilon$, where the distribution of size does not change strongly from the initial one with small oscillations around the average value (Poissonian distribution). In this case, in fact, the opinion dynamics process is much faster than the dynamics of the network, leading easily to an equilibrium state before the groups have the possibility to re-shape. In the following, we will explore in more details the case with $\varepsilon=0.1$. 

\begin{figure}[!h]
  \includegraphics[width=12cm,clip=true]{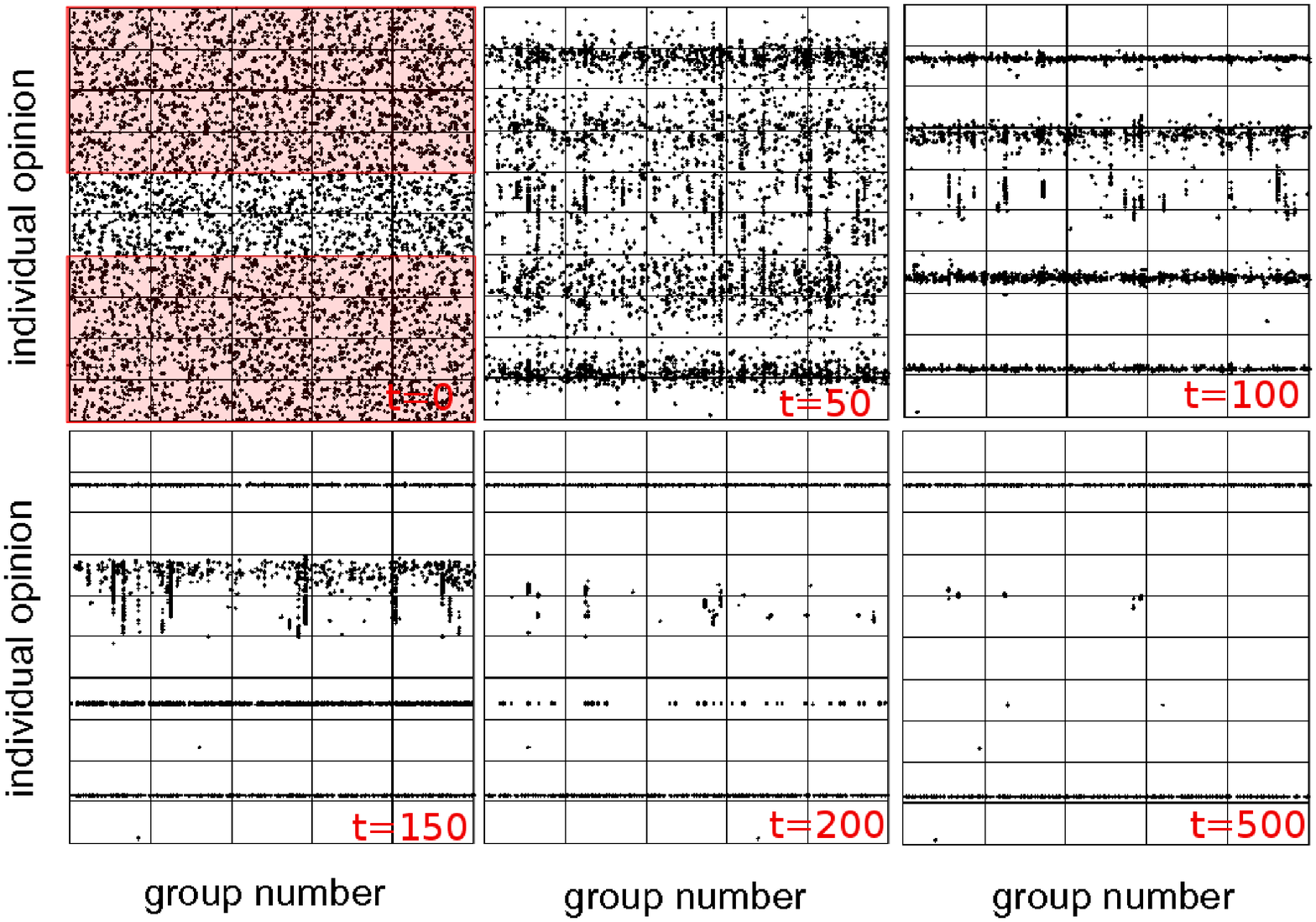}
   \includegraphics[width=12cm,clip=true]{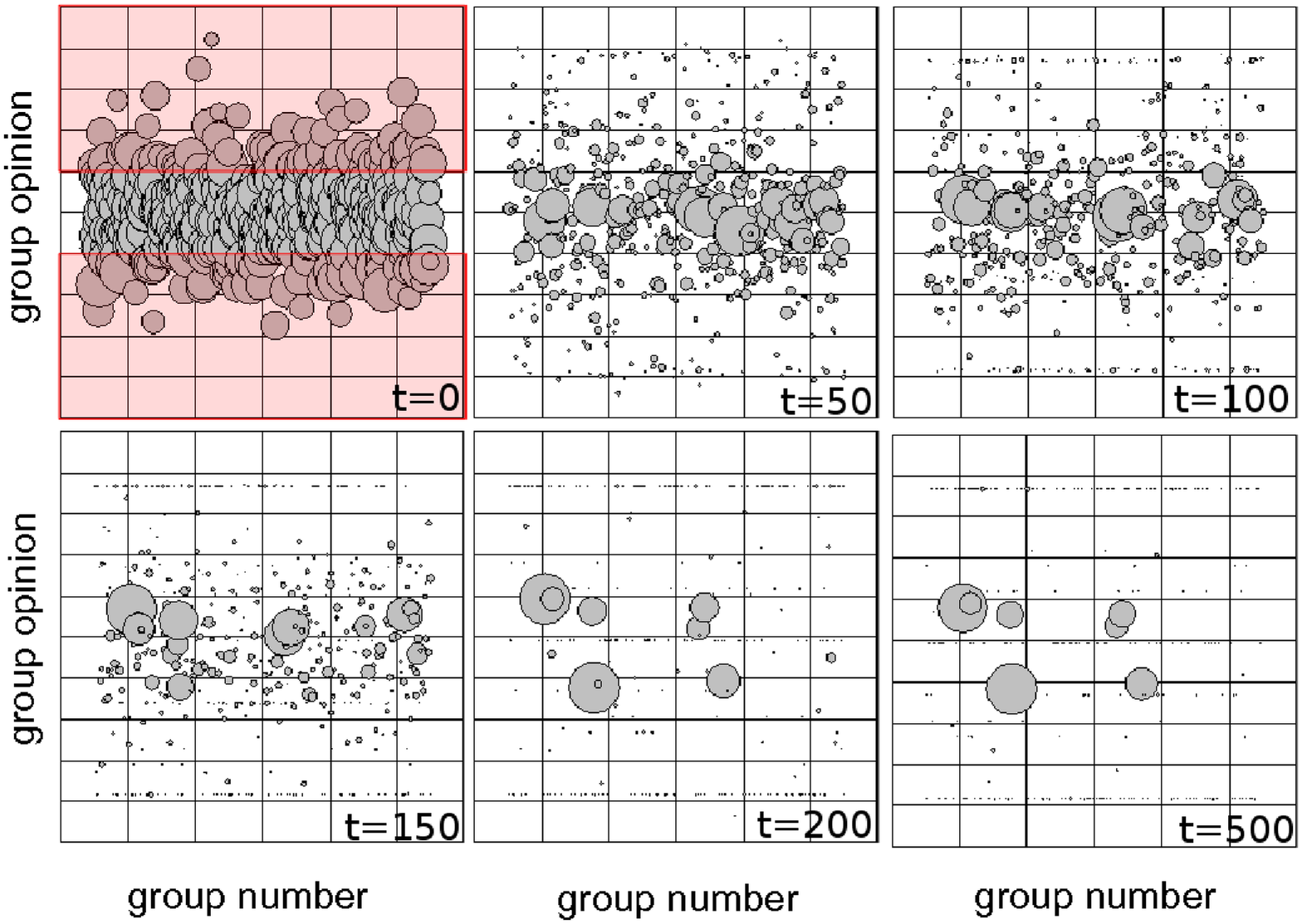}
  \caption{Evolution of a population in one realization for $p_{change}=0.5$ and $\varepsilon=0.1$ (from the left to the right, after 0, 50, 100, 150, 200 and 500 iterations)
. Upper plots: individual opinions. Lower plots: average group opinions and relative group size (radius of the ball) }\label{fig_1real_BCgroups01}
\end{figure}

The figure \ref{fig_1real_BCgroups01} represents the evolutions, in one realization of the system, of individuals (upper plots) and groups (lower plots) for $\varepsilon=0.1$. We can identify two different stages: a first one where we observe the formation of many extremist formations (t = 50 for group graph) and a second one where the central formations tend to concentrate in few larger groups (t$\geq$200). From the figure we can also observe that the formation of opinion clusters happens at a very early stage of the evolution. 

The first stage, namely the radicalization of groups, is due to the fact that, at the beginning all the agents in the shadowed area ($<\vartheta_i><0.5-\varepsilon$ and $<\vartheta_i>>0.5+\varepsilon$) have a very high probability to be unsatisfied by their group and therefore to change group. These agents start to move to the groups which, due to the random initialization (or the group change dynamic by itself), already have an opinion slightly drifted to their extreme, shifting therefore the  opinion of their new group more and more in their direction. The opposite drift happens on the old group that consequently move toward the other extreme. The movement of these agents,  generates an imbalance in the average group opinions that gradually assume a wide spectrum (t = 0 to 100 from group graphs).  In the mean time the opinion dynamics is concluded: the more extremist groups contain a single cluster with extremist opinion and, therefore, they are very stable. But all the other groups still contain more that one opinion cluster. In particular some central groups started to enlarge at this stage, being able to capture, from the both sides, the more central agents excluded by the radicalization of the groups. This is the starting point of the preferential attachment.

It is now the turn of moving groups for the agents with a more central opinion: they are no more satisfied by the groups that maintained an opinion ($O\sim 0.5$) and try to search groups more affine to them. The larger groups are favoured to be selected as the new group, and the movement of the agents leads gradually these groups to the position of the central clusters (t = 200 for group graph). Once this position is reached, these larger and moderate groups are able to capture all the agents, with the same opinion of the group, that are still grouped with more extremist agents. At the end almost all the groups contain a single cluster; but, since the mechanism of formation of the extremist groups is different from the one of the central ones, the sizes of these structures are different: extremist groups are usually very numerous and composed of few individuals while moderate groups are few and very large. That is what we observe, at the t=500 in Figure \ref{fig_1real_BCgroups01}.

In the following we will show the aggregate results about the distributions of opinions on 100 replicas of the simulations.

\begin{center}
\begin{figure}[!h]
  \includegraphics[width=10cm,clip=true]{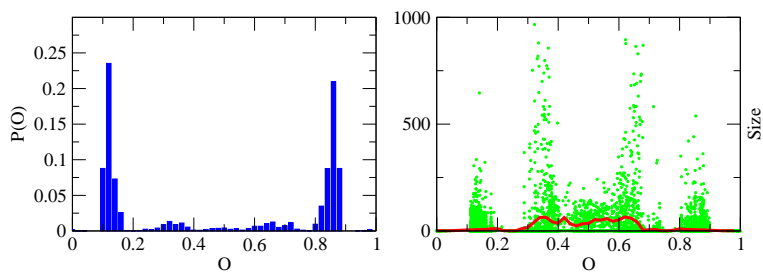}
  \caption{Left plot:Opinion distribution  for groups for $\varepsilon=0.1$. Right plot: Group opinion versus group size for $\varepsilon=0.1$. The red line represents the average size, for a given group opinion.  The result is averaged on 100 replicas of the simulation.}\label{fig_pop}
\end{figure}
\end{center}

In the left plot of figure \ref{fig_pop} the average opinion distribution $P(O)$ of the groups is represented, for $\varepsilon=0.1$. The distribution presents a clear bimodal behaviour picked on quite extreme opinions. Such a distribution means that the majority of the groups are extremist formations, characterized by an opinion strongly different from the neutral one. 
In the right plot of \ref{fig_pop} it is described the relation between the size of a group and its opinion. As we can notice from the correlation between group sizes and their relative opinions, to extremist opinions (that are the majority for the groups) correspond, on average, the smallest group sizes. Therefore, the small fraction of macroscopic groups appearing in the scale free size distribution of figure \ref{fig_psize} represents mostly the "moderate" group formations. 

\begin{center}
\begin{figure}[!h]
  \includegraphics[width=10cm,clip=true]{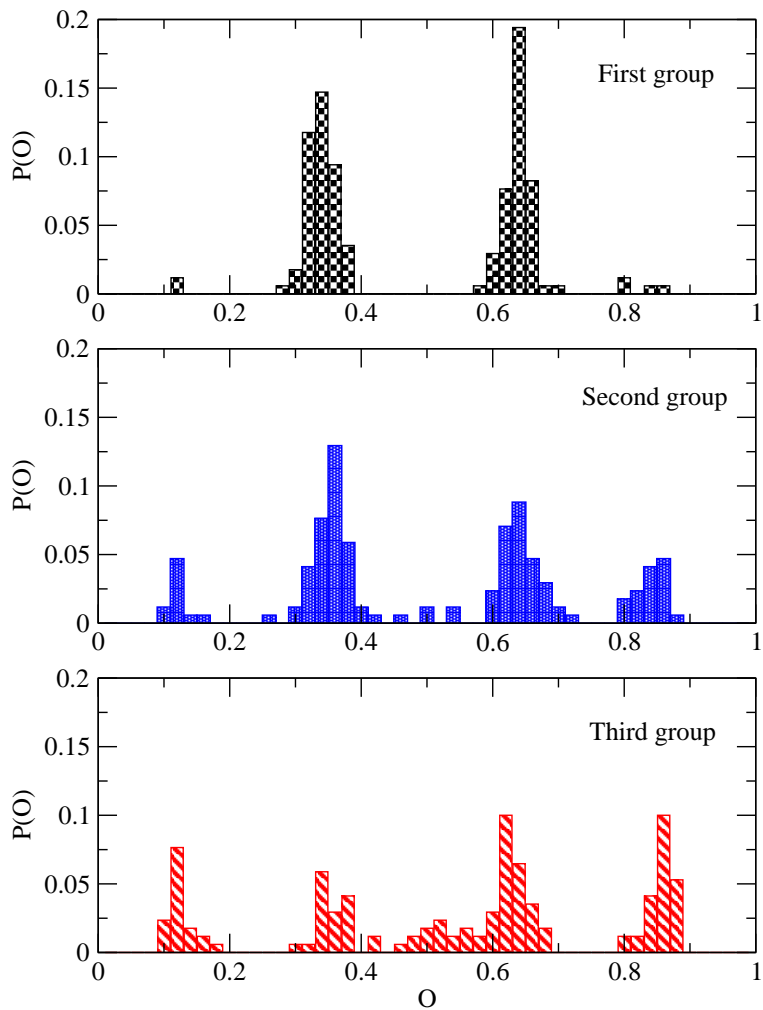}
  \caption{First (black), second (red), third (blue) group opinion distribution for $\varepsilon=0.1$. The result represents  500 replicas of the simulation.}\label{fig_unduetre}
\end{figure}
\end{center}

Also, as we can observe from the scattered points in Figure \ref{fig_pop}, some very large extremist formations are also present in the scenario. Sometimes, though rarely, it can also happen that an extremist formation is the biggest group. This rare situation can be realized if a group, has an average extreme opinion since the beginning and therefore it results able to attract extremist agents already in the first stage (radicalization of groups) of the evolution. In figure \ref{fig_unduetre}  the opinion distribution for the biggest group, the second group and the third is represented. We can observe that with high probability the biggest group is one of the few moderate formations present in the scenario. But with a lower probability it can also be an extremist formation. For the second and the third group, the probability becomes more and more uniform in the opinion space. 

To conclude and to go back to some considerations regarding the opinion structure inside the network, it is important to correlate the opinion cluster structures and the group structures defined for this particular case. The question that can arise is: are the group structures also opinion clusters?

To answer this question we should first of all define what we mean by "opinion clusters". The definition we used in the previous section is the one commonly used in most of the papers on the topic. But it is difficult to apply such a definition to answer our question regarding groups. In the following we reformulate the question in the following way: how is the opinion spread inside the topological groups? Fixing a threshold $\Delta\vartheta_{MAX}=0.05$, we say that one group is an opinion cluster if: $\vartheta_{MAX}-\vartheta_{min}<\Delta\vartheta_{MAX}$. Using this definition we can observe that the number of groups that are also opinion cluster (for $\varepsilon=0.1$) is very high: $\%_{clust}=0.976$. It means that in most cases, a group contains only one opinion cluster.

\section{Conclusions}

We proposed a model based on a heuristic inspired from \cite{kozma2008consensus}. However, the decision for an individual to change its social network is not linked to its distance to one of its neighbour. On the contrary, it is linked to its distance to its own group opinion. A too large distance means a discomfort or a disagreement with its own group and push the individual to change groups. More concretely, an individual, with a given probability, become closer from the opinion point of view to one of its neighbour if it is already close enough. It can, with the complementary probability, change groups if it is far from the average opinion of the individuals of its own group.  

The model described in this paper is quite complex and presents a high dimensional parameter space. Here, a small portion of the parameter space has been explored, keeping most of them fixed. 

The simulations in the explored area of the parameter space already show very interesting behaviours regarding the co-evolution of group structures and opinions. First of all, the presence of a co-evolving group based, network structure changes the dynamical behaviour of opinion dynamics process. We analyzed a continuous opinion dynamics evolution mechanism (Deffuant model) that shows a phase transition between pluralistic and consensus scenarios. The phase parameter for such transition is the tolerance parameter, describing how close, from the point of view of the opinion, two persons must be, to have a discussion making them closer to each other. The transition properties are the same on all static network topologies. 

However, and it is our first interesting result, we can clearly observe that the transition happens at lower values of the phase parameter in our case of co-evolving network structure. It means that the double process of dynamics on network and of dynamics of the network favours the consensus formation. The network structure is intimately linked to the group approach. In the model, in the absence of groups into the population, some extremist minor clusters remain (\cite{ben2003bifurcations}) and make the consensus threshold of tolerance quite high. Comparing our work to the model presented in \cite{kozma2008consensus}, linking the decision of social network change to the group distance instead of only a neighbour distance, allows to obtain a complete inverse conclusion. Indeed, while \cite{kozma2008consensus} conclude that rewiring increases the number of final opinion clusters for large tolerance, we argue that it leads to the consensus for a smaller tolerance value than the classical one required by the Deffuant Bounded Confidence Model. With this group approach, our model is finally closer to the one of \cite{hegselmann2002opinion} which also obtain the consensus with a lower threshold of tolerance. To obtain this result, it considers that the next opinion adopted by the individual is the average of the opinion of all individuals having a sufficiently close opinion to its own. From our model, we conclude than when there is the possibility to change groups if you disagree with your own, no extremist minor clusters exists and the consensus is reached for a lower level of tolerance. One can notice in the real life that the membership avoids isolation and favours a high degree of cohesion. Our model seems to be also able to avoid isolation by the membership mechanism.

Moreover, in the model, a relatively large tolerance on the discussed subject implies that the group existence and social weight is not affected by the discussion. Indeed the distribution of group sizes does not widely changes for large tolerance values. Looking around us, we can observe that, some times, the open-minded groups can debate on various subjects without challenging the group itself.

The situation is different when the tolerance is lower. That is the second interesting behaviour we notice from our first study. For a case in the pluralistic region where the tolerance is lower, we observed a power law group size distribution, corresponding to a final state with strong heterogeneity between group sizes.  We also studied the correlation between such distribution and the opinion distribution for groups. We observed that most of the groups formed are extremist formations but the small minority of central groups is, on average, the biggest in terms of size.

Indeed, for a tolerance of 0.1, a majority of people tends to group with each other in few groups that generally adopt an almost centred position. Also, a lot of small extremist groups remain. At the population level, the number of opinion clusters remains the same as if there were no group in the population. 

For this particular level of tolerance, the group equilibrium is changed by the discussion. As in real life, when people are highly involved in a discussed subject, they can change groups just because they disagree. When individuals are not tolerant about something, they can be comfortable in a group only if this group is cohesive enough. That implies not containing people having very different opinion on this subject. Thus, unexpectedly, in the model, this intolerance leads to the emergence of few centred major group with quite a cohesive opinion. That looks like the political situation in many countries where individuals join a particular party not for a real ideological choice but expecting the party to act on some issue particularly important for them, for example the environment. Moreover, the heterogeneous characteristic of the distribution of size of the groups seems particularly realistic even if it is not the only distribution form we can observe. Looking at the example of the size of political parties, the typical situation in many republics is that there are two big parties, and a relatively large number of very small ones. That is the example for France. There are however interesting (relatively stable) situations of single dominant parties (Venezuela now, for example), and others of just a multitude of minorities (Russia in the first years after the break-down of the USSR, for example). 

A deeper analysis, centered on the group hierarchy, for different values of the tolerance parameter $\varepsilon$, is given in \cite{Gargiulo:ECCS}, where it is also analyzed the stability of the group structures. The bifurcation plot regarding the details of the opinion dynamics process, instead is presented in \cite{Gargiulo:WEIN}. 

The behaviour of the model is also strongly influenced by the number of groups, $G_0$ and $N$ the size of the population. Clearly, if $G_0 = 1$ or $G_0 = N$  the usual BC is obtained, but the spectrum of behaviours between these two values is quite wide. We know, for example, that the number of initial members in a group, in fact, allows the opinion dynamics process to start or not inside a group (\cite{Gargiulo:ECCS}). The analysis of the behaviours according to these parameters is quite complex and deserves indeed an analytical study in a simplified contest. This analysis goes beyond the aim of the present paper and will be developed in  a forthcoming work.

\section*{Acknowledgments}
The authors would like to thank Santo Fortunato and Guillaume Deffuant for the useful discussions. We also thank our referees for their interesting comments which have helped us to improve the paper.
\section*{References}
\bibliographystyle{plain}	
\bibliography{GO_refs}

\begin{thebibliography}{10}

\bibitem{ard2004role}
F.~Amblard and G.~Deffuant.
\newblock {The role of network topology on extremism propagation with the
  relative agreement opinion dynamics}.
\newblock {\em Physica A: Statistical Mechanics and its Applications},
  343:725--738, 2004.

\bibitem{ben2003bifurcations}
E.~Ben-Naim, PL~Krapivsky, and S.~Redner.
\newblock {Bifurcations and patterns in compromise processes}.
\newblock {\em Physica D: Nonlinear Phenomena}, 183(3-4):190--204, 2003.

\bibitem{Byrne1971}
D.~Byrne.
\newblock {The Attraction Paradigm}.
\newblock 1971.

\bibitem{castellano2009statistical}
C.~Castellano, S.~Fortunato, and V.~Loreto.
\newblock {Statistical physics of social dynamics}.
\newblock {\em Reviews of Modern Physics}, 81(2):591--646, 2009.

\bibitem{clifford1973model}
P.~Clifford and A.~Sudbury.
\newblock {A model for spatial conflict}.
\newblock {\em Biometrika}, 60(3):581, 1973.

\bibitem{deffuant2000mixing}
G.~Deffuant, D.~Neau, F.~Amblard, and G.~Weisbuch.
\newblock {Mixing beliefs among interacting agents}.
\newblock {\em Advances in Complex Systems}, 3(4):87--98, 2000.

\bibitem{Gargiulo:ECCS}
Gargiulo F. and S.~Huet.
\newblock {How opinion dynamics generates group hierarchies}.
\newblock {\em arXiv/1003.3560}, 2010.

\bibitem{Gargiulo:WEIN}
Gargiulo F. and S.~Huet.
\newblock {When group level is different from the population level: an adaptive
  network with the Deffuant model}.
\newblock {\em arXiv/1002.1896}, 2010.

\bibitem{festinger1957Dissonance}
L.~Festinger.
\newblock {A Theory of Cognitive Dissonance}.
\newblock page 291, 1957.

\bibitem{fortunato2004universality}
S.~Fortunato.
\newblock {Universality of the Threshold for Complete Consensus for the Opinion
  Dynamics of Deffuant et al.}
\newblock {\em International Journal of Modern Physics C}, 15:1301--1307, 2004.

\bibitem{2009community}
S.~Fortunato.
\newblock {Community detection in graphs}.
\newblock {\em Physics Reports}, 486:75--174, 2010.

\bibitem{galam2008sociophysics}
S.~Galam.
\newblock {Sociophysics: A review of Galam models}.
\newblock {\em International Journal of Modern Physics C}, 19(3):409--440,
  2008.

\bibitem{gargiulo2008can}
F.~Gargiulo and A.~Mazzoni.
\newblock {Can extremism guarantee pluralism?}
\newblock {\em JASSS}, 11(4).

\bibitem{girvan2002community}
M.~Girvan and MEJ Newman.
\newblock {Community structure in social and biological networks}.
\newblock {\em Proceedings of the National Academy of Sciences}, 99(12):7821,
  2002.

\bibitem{gross2008adaptive}
T.~Gross and B.~Blasius.
\newblock {Adaptive coevolutionary networks: a review}.
\newblock {\em Journal of the Royal Society Interface}, 5(20):259, 2008.

\bibitem{hegselmann2002opinion}
R.~Hegselmann and U.~Krause.
\newblock {Opinion Dynamics and Bounded Confidence Models, Analysis and
  Simulation}.
\newblock {\em Journal of Artificial Societies and Social Simulation}, 5, 2002.

\bibitem{heider1946}
F.~Heider.
\newblock {Attitudes and Cognitive Organization}.
\newblock {\em The Journal of Psychology}, 21:107--112, 1946.

\bibitem{helbing2000simulating}
D.~Helbing, I.~Farkas, and T.~Vicsek.
\newblock {Simulating dynamical features of escape panic}.
\newblock {\em Nature}, 407(6803):487--490, 2000.

\bibitem{holley1978survival}
R.~Holley and TM~Liggett.
\newblock {\em The Annals of Probability}, 3.

\bibitem{hu2009wifi}
H.~Hu, S.~Myers, V.~Colizza, and A.~Vespignani.
\newblock {WiFi networks and malware epidemiology}.
\newblock {\em Proceedings of the National Academy of Sciences}, 106(5):1318,
  2009.

\bibitem{huet2008rejection}
S.~Huet, G.~Deffuant, and W.~Jager.
\newblock {A Rejection Mechanism In 2d Bounded Confidence Provides More
  Conformity}.
\newblock {\em Advances in Complex Systems}, 11(4):529--549, 2008.

\bibitem{kozma2008consensus}
B.~Kozma and A.~Barrat.
\newblock {Consensus formation on adaptive networks}.
\newblock {\em Physical Review E}, 77(1):16102, 2008.

\bibitem{laguna2004minorities}
MF~Laguna, G.~Abramson, and D.H. Zanette.
\newblock {Minorities in a model for opinion formation}.
\newblock {\em Complexity}, 9(4):31--36, 2004.

\bibitem{lorenz2007continuous}
J.~Lorenz.
\newblock {Continuous Opinion Dynamics Under Bounded Confidence:. a Survey}.
\newblock {\em International Journal of Modern Physics C}, 18:1819--1838, 2007.

\bibitem{MatzWood2005}
D.~Matz and W.~Woody.
\newblock {Cognitive dissonance in groups: The consequences of disagreement}.
\newblock {\em Journal of personality and social psychology}, 88(1):22--37,
  2005.

\bibitem{palla2007quantifying}
G.~Palla, A.L. Barab{\'a}si, and T.~Vicsek.
\newblock {Quantifying social group evolution}.
\newblock {\em NATURE-LONDON-}, 446(7136):664, 2007.

\bibitem{pastor2001epidemic}
R.~Pastor-Satorras and A.~Vespignani.
\newblock {Epidemic spreading in scale-free networks}.
\newblock {\em Physical review letters}, 86(14):3200--3203, 2001.

\bibitem{pastor2002immunization}
R.~Pastor-Satorras and A.~Vespignani.
\newblock {Immunization of complex networks}.
\newblock {\em Physical Review E}, 65(3):36104, 2002.

\bibitem{pastor2004evolution}
R.~Pastor-Satorras and A.~Vespignani.
\newblock {\em {Evolution and structure of the Internet: A statistical physics
  approach}}.
\newblock Cambridge Univ Pr, 2004.

\bibitem{radicchi2004defining}
F.~Radicchi, C.~Castellano, F.~Cecconi, V.~Loreto, and D.~Parisi.
\newblock {Defining and identifying communities in networks}.
\newblock {\em Proceedings of the National Academy of Sciences}, 101(9):2658,
  2004.

\bibitem{SherifHovland1961}
M.~Sherif and C.~I. Hovland.
\newblock {Social judgment: Assimilation and contrast effects in communication
  and attitude change}.
\newblock 1961.

\bibitem{stauffer2002sociophysics}
D.~Stauffer.
\newblock {Sociophysics: the Sznajd model and its applications}.
\newblock {\em Computer physics communications}, 146(1):93--98, 2002.

\end{thebibliography}
\end{document}